\documentclass{aa}
\usepackage{graphics}
\begin{document}

\thesaurus{01     
              (07.16.2;  
               13.09.5)}  

   \title{Jupiter's hydrocarbons observed with ISO-SWS:\\
vertical profiles of $\rm C_2H_6$ and $\rm C_2H_2$, detection of $\rm CH_3C_2H$}

   \author{T.\ Fouchet\inst{1}
          \and
          E.\ Lellouch\inst{1}
	  \and
	  B.\ B\'ezard\inst{1}
	  \and
          H.\ Feuchtgruber\inst{2}
	  \and
	  P.\ Drossart\inst{1}
	  \and
	  T.\ Encrenaz\inst{1}
          }

   \institute{DESPA, Observatoire de Paris, 5 Place Jules Janssen,
              92195 Meudon Cedex, France\\
              email: thierry.fouchet@obspm.fr
         \and
             Max-Planck-Institut f\"ur Extraterrestrische Physik,
             85748 Garching, Germany}

   \offprints{T.\ Fouchet}

   \date{Received; accepted}

   \titlerunning{Jupiter's hydrocarbons observed with ISO-SWS}
   \maketitle

\begin{abstract}

We have analysed the ISO-SWS spectrum of Jupiter in the 12--16 $\mu{\rm m}$
range, where several hydrocarbons exhibit rovibrational bands. Using
temperature information from the methane and hydrogen emissions, we derive the
mixing ratios ($q$) of acetylene and ethane at two independent pressure levels.
For acetylene, we find $q=(8.9^{+1.1}_{-0.6})\times10^{-7}$ at 0.3 mbar and
$q=(1.1^{+0.2}_{-0.1})\times10^{-7}$ at 4 mbar, giving a slope
$-d\ln q / d\ln P=0.8\pm0.1$, while for ethane $q=(1.0\pm0.2)\times10^{-5}$ at
1 mbar and $q=(2.6^{+0.5}_{-0.6})\times10^{-6}$ at 10 mbar, giving
$-d\ln q / d\ln P=0.6\pm0.2$. The ethane slope is consistent with the
predictions of Gladstone et al.\ (\cite{Gladstone96}), but that predicted for
acetylene is larger than we observe. This disagreement is best explained by an
overestimation of the acetylene production rate compared to that of ethane in
the Gladstone et al.\ (\cite{Gladstone96}) model.

At 15.8 $\mu{\rm m}$, methylacetylene is detected for the first time at low
jovian latitudes, and a stratospheric column density of
$(1.5\pm0.4)\times10^{15}$ molecule\,cm$^{-2}$ is inferred. We also derive an
upper limit for the diacetylene column density of $7\times10^{13}$
molecule\,cm$^{-2}$.


\end{abstract}

\section{Introduction}

\indent\indent Hydrocarbons in Jupiter are produced in a series of chemical pathways initiated by the photolysis of methane in the upper stratosphere. Vertical transport, mainly turbulent diffusion, redistributes the molecules throughout the stratosphere and down to the troposphere, where they are eventually destroyed. Hydrocarbons, and particularly the most stable of them, are therefore good tracers of the upper atmospheric dynamics. In addition, as they strongly contribute to the atmospheric opacity in the UV and IR, hydrocarbons act as major sources and sinks of heat, thereby participating to the stratospheric dynamics.

All these reasons have strongly motivated theoretical studies of the jovian stratospheric photochemistry (Strobel \cite{Strobel69}; Yung \& Strobel \cite{Yung80}; and most recently Gladstone et al.\ \cite{Gladstone96}). Although nowadays very detailed, these models still need to be constrained by observations of minor species. However, prior to the ISO mission, only two molecules, acetylene (${\rm C_2H_2}$) and ethane (${\rm C_2H_6}$), had been detected, except in the auroral zones where several other minor species (${\rm C_2H_4}$, ${\rm C_3H_4}$ and ${\rm C_6H_6}$) have been observed (Kim et al.\ \cite{Kim85}). Mean stratospheric mole fractions have been inferred for ${\rm C_2H_2}$ and ${\rm C_2H_6}$ by various authors, but no precise information on their vertical variations was made available. In this paper, we analyse the ISO-SWS spectrum of Jupiter between 7 and 17 $\mu{\rm m}$, in order to determine the vertical distributions of ${\rm C_2H_2}$ and ${\rm C_2H_6}$, and to search for more complex (C$_3$ and C$_4$) molecules. Sect.~2 presents the observations. Our analysis of the spectrum is presented in Sect.~3. The results are compared with previous observations and theoretical predictions in Sect.~4.


\section{Observations}

\indent\indent Descriptions of the Infrared Space Observatory (ISO) and of the Short Wavelength Spectrometer (SWS) can be found respectively in Kessler et al.\ (\cite{Kessler96}) and de Graauw et al.\ (\cite{deGraauw96}). A preliminary analysis of the Jupiter SWS spectrum can be found in Encrenaz et al.\ (\cite{Encrenaz96}). New ISO-SWS grating observations of Jupiter were obtained on May 23, 1997 UT using the AOT 01 observing mode. These observations have an average spectral resolution of 1500, and range from 2.4 to 45 $\mu{\rm m}$. However, the useful range is limited to 2.4--17 $\mu{\rm m}$, due to partial saturation at longer wavelengths. The instrument aperture, $14\times20$ arcsec$^2$ at $\lambda<12.5$ $\mu{\rm m}$ and $14\times27$ arcsec$^2$ at $\lambda>12.5$ $\mu{\rm m}$, was centered on the planet with the long axis aligned perpendicular to the ecliptic, thus roughly parallel to the N-S polar axis. It covered latitudes between 30$^{\circ}$S and 30$^{\circ}$N, and $\pm20^{\circ}$ longitude range from the central meridian. The absolute flux accuracy is $\sim20\%$.

Instrumental fringing generates a spurious signal between 12.5 and 17 $\mu{\rm m}$, which amounts to $\sim$10\%
of the continuum level. This parasitic signal was for the most part removed by fitting the detector relative response function to the observed spectrum and then dividing it out. Residual fringes were further removed by selective frequency filtering.


\section{Analysis}

\begin{figure}
 \resizebox{\hsize}{!}{\includegraphics{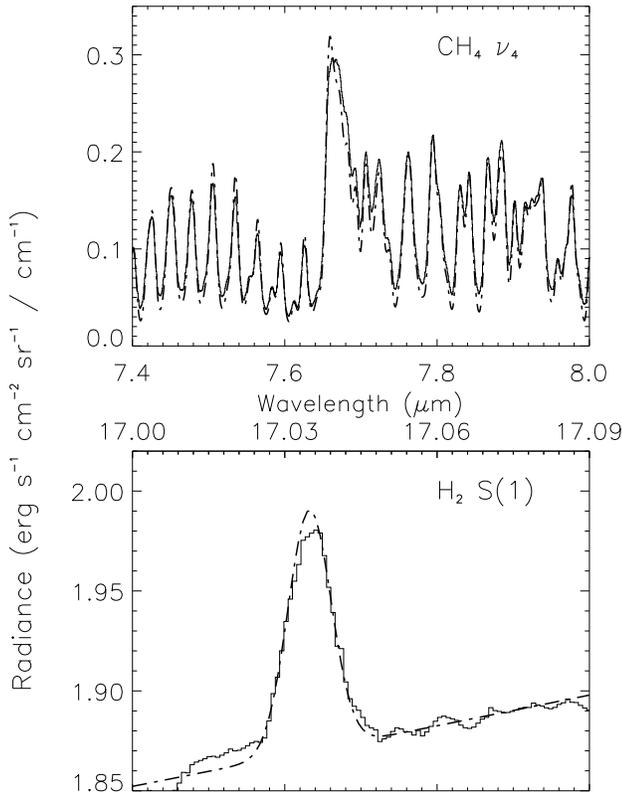}}
 \caption{Comparison between ISO-SWS spectra (solid line) and synthetic spectra (dotted line) in the ${\rm CH_4}$ $\nu_4$ band (upper panel) and the ${\rm H_2}$ S(1) line (lower panel)}
\label{FigTemp}
\end{figure}

\indent\indent We analysed the ISO-SWS spectrum using a standard line-by-line radiative transfer code adapted to Jupiter's conditions. We included the molecular absorptions by ${\rm NH_3}$, ${\rm CH_4}$, ${\rm C_2H_2}$, ${\rm C_2H_6}$, and ${\rm CH_3C_2H}$ using the spectroscopic parameters of the GEISA97 databank (Jacquinet-Husson et al.\ \cite{Husson99}). We also considered ${\rm C_4H_2}$ absorption, using a linelist provided by E.\ Ari\'e (private communication) and band intensities from Koops et al.\ (\cite{Koops84}). Spectroscopic parameters for the ${\rm H_2}$ S(1) line were calculated using molecular constants from Jennings et al.\ (\cite{Jennings87}) and Reuter \& Sirota (\cite{Reuter94}). The H$_2$-He collision-induced continuum was calculated following the work of Borysow et al.\ (\cite{Borysow85}, \cite{Borysow88}). The ${\rm NH_3}$ vertical distribution was taken from Fouchet et al. (\cite{Fouchet99}).

\subsection{Temperature profile}

\indent\indent We first calculated synthetic spectra in the region of the ${\rm CH_4}$ $\nu_4$ band. For ${\rm CH_4}$, we used a deep mixing ratio of $2.1\times10^{-3}$ (Niemann et al.\ \cite{Niemann98}) and the vertical profile derived by Drossart et al.\ (\cite{Drossart99}) from the ${\rm CH_4}$ fluorescence emission at 3.3 $\mu{\rm m}$. The $\nu_4$ band allows one to retrieve 4 independent points on the temperature profile between 35 and 1 mbar. We also generated synthetic spectra of the ${\rm H_2}$ S(1) rotational line at 17 $\mu{\rm m}$. This line probes a broad atmospheric layer at 3--30 mbar. The ortho-to-para ratio of ${\rm H_2}$ was assumed to follow local thermodynamical equilibrium. The stratospheric temperature profile was adjusted in order to best match the absolute emission in the ${\rm CH_4}$ band, and the line-to-continuum ratio of the ${\rm H_2}$ S(1) line.

\begin{figure}
 \resizebox{\hsize}{!}{\includegraphics{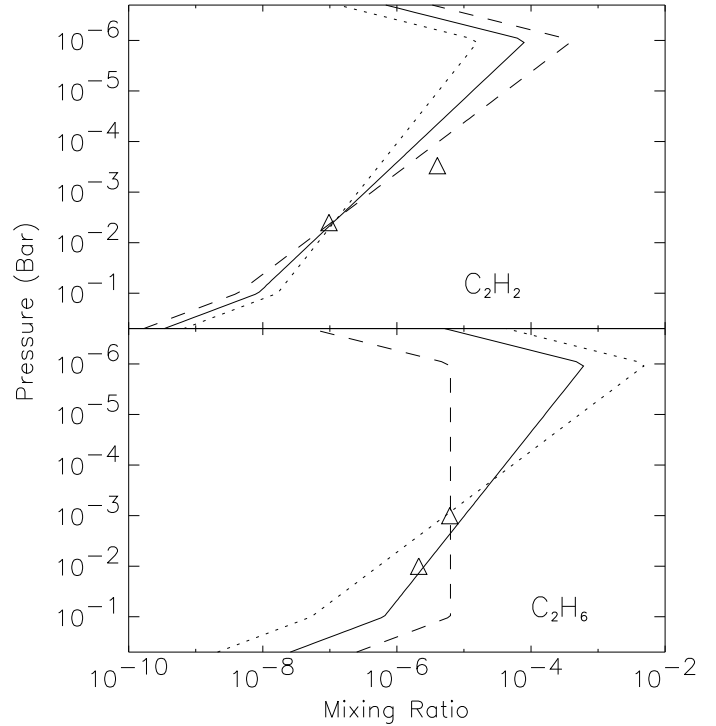}}
 \caption{Vertical profiles of ${\rm C_2H_2}$ (upper panel) and ${\rm C_2H_6}$ (lower panel) used for the calculation of the synthetic spectra of Fig.~\ref{FigAce} and Fig.~\ref{FigEth}. The best-fit profiles are shown as solid lines. The triangles are from the Gladstone et al.\ (\cite{Gladstone96}) model.}
\label{FigPro}
\end{figure}

In practice, starting with the temperature profile measured in-situ by the Galileo Probe (Seiff et al.\ \cite{Seiff98}), it was necessary to cool it by 2 K between 30 and 5 mbar. At pressures lower than 5 mbar, the initial profile was warmed by 4 K up to $165$ K. The $\nu_4$ band is also somewhat sensitive to the temperature around the 10-$\mu$bar pressure level. We found that the temperature remains constant within a few degrees between 1 mbar and 1 $\mu$bar, as already observed by Seiff et al.\ (\cite{Seiff98}). At pressures lower than 1 $\mu$bar, we adopted the measurements of Seiff et al.\ (\cite{Seiff98}), vertically smoothed in order to remove oscillations due to gravity waves, noting that our measurements are essentially insensitive to this pressure range. The fit to the ${\rm CH_4}$ and ${\rm H_2}$ emissions is presented in Fig.~\ref{FigTemp}.

The 20\%
uncertainty on the absolute flux calibration directly results in an uncertainty of $\pm2 $K on the temperature profile inferred from the ${\rm CH_4}$ emission. This uncertainty partly explains the minor disagreement between the modelled and observed ${\rm H_2}$ line. Indeed, our model, while giving an optimum fit to the ${\rm CH_4}$ emission, slightly (5--10\%) overpredicts the observed line-to-continuum ratio of the S(1) line. We also note that the ${\rm H_2}$ ortho-to-para ratio could differ from the thermal equilibrium value, as observed in the troposphere by Conrath et al. (\cite{Conrath98}). For example, a synthetic spectrum calculated with a constant para fraction of 0.34, corresponding to the thermal equilibrium value at 115 K, would fully reconcile the ${\rm CH_4}$ and ${\rm H_2}$ measurements.
\subsection{Acetylene}

\begin{figure}
 \resizebox{\hsize}{!}{\includegraphics{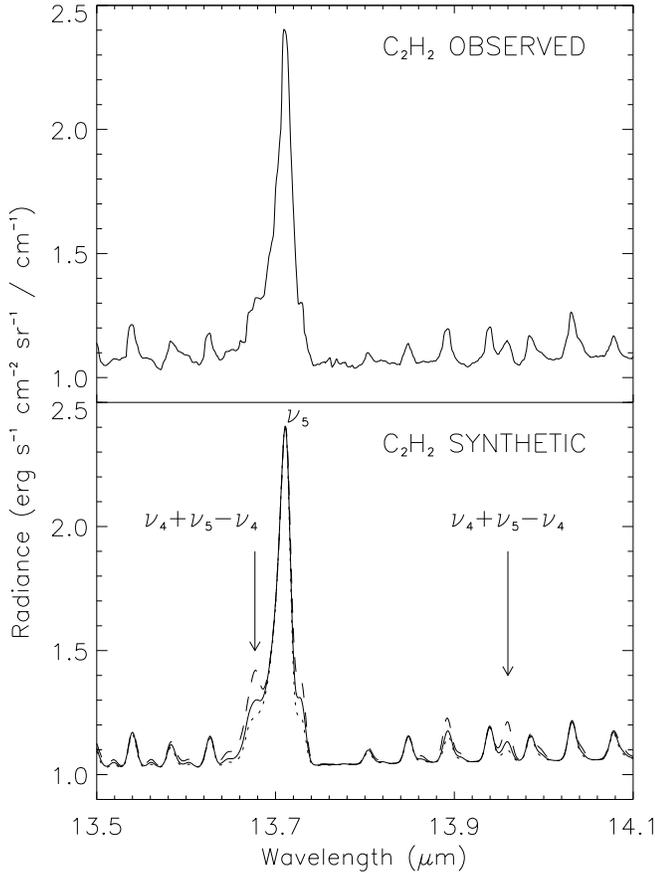}}
 \caption{Comparison near 13.7 $\mu{\rm m}$ between the ISO-SWS spectrum (upper panel) and three synthetic spectra (lower panel) calculated with the ${\rm C_2H_2}$ vertical distributions of Fig.~\ref{FigPro}. Solid line gives the best fit.}
 \label{FigAce}
\end{figure}

\indent\indent The ISO-SWS observations in the vicinity of the ${\rm C_2H_2}$ emission at 13.7 $\mu{\rm m}$ were compared with synthetic spectra obtained with three distinct ${\rm C_2H_2}$ vertical distributions (Fig.~\ref{FigPro}). All three profiles reproduce the emissions due to the P-,Q-, and R-branches of the main $\nu_5$ ${\rm C_2H_2}$ band (Fig.~\ref{FigAce}). However, only one profile (Fig.~\ref{FigPro}, solid line) fits the observed spectrum in the Q-branches of the $\nu_4+\nu_5-\nu_4$ band at 13.68 and 13.96 $\mu{\rm m}$. Indeed, while the $\nu_5$ band probes atmospheric levels between 2 and 5 mbar, the hot band sounds warmer, higher levels around 0.3 mbar, allowing us to determine the slope of the ${\rm C_2H_2}$ profile between these two regions.

Error bars on the ${\rm C_2H_2}$ mixing ratio were estimated by taking into account instrumental noise and the uncertainty in the relative strengths of the ${\rm C_2H_2}$ lines. The resulting mixing ratios are $q=(8.9^{+1.1}_{-0.6})\times10^{-7}$ at 0.3 mbar and $q=(1.1^{+0.2}_{-0.1})\times10^{-7}$ at 4 mbar, giving a slope $-d\ln q / d\ln P=0.8\pm0.1$. The error on the temperature profile ($\pm2 $K) introduces an additional uncertainty on the ${\rm C_2H_2}$ mixing ratios of about 20\%.
This error, however, essentially equally applies to all pressure levels, and thus leaves the retrieved ${\rm C_2H_2}$ profile slope mostly unaffected.

\subsection{Ethane}

\begin{figure}
 \resizebox{\hsize}{!}{\includegraphics{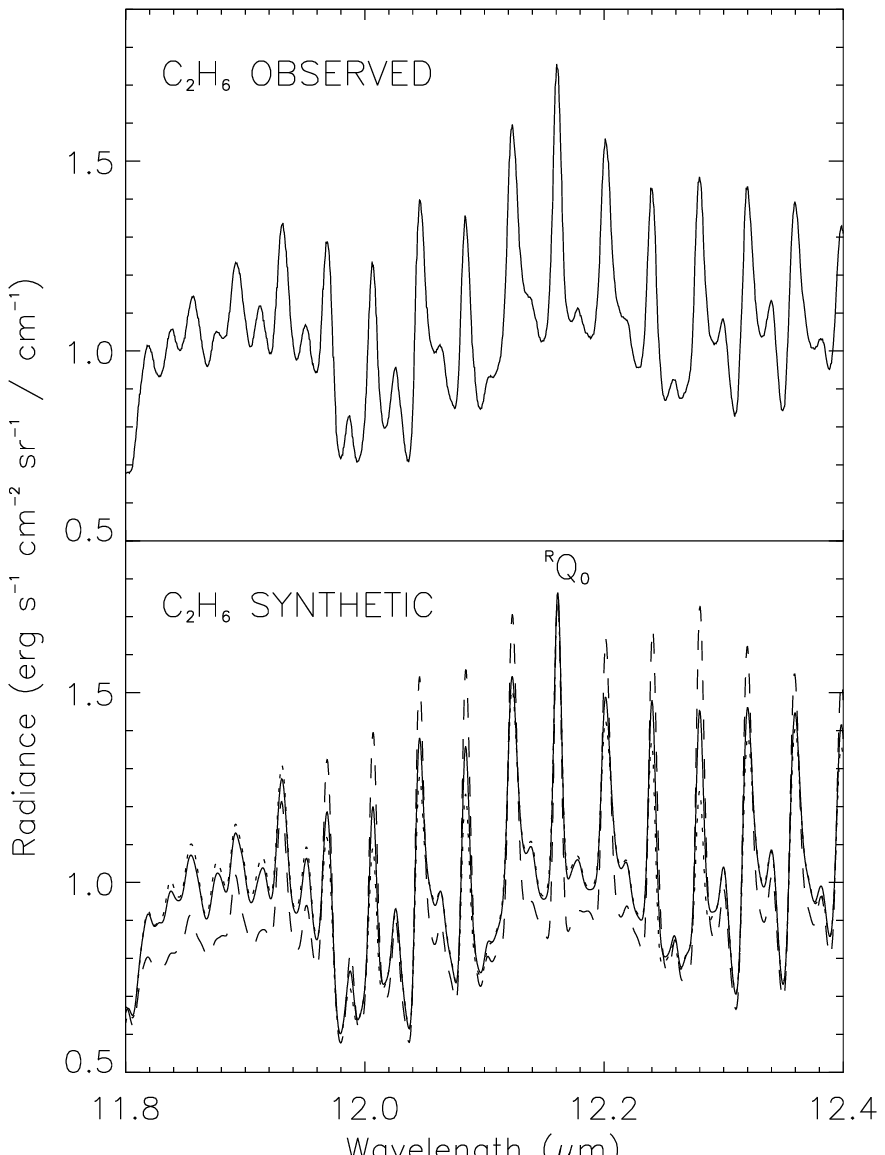}}
 \caption{Same as Fig.~\ref{FigAce} for ${\rm C_2H_6}$ at 12 $\mu{\rm m}$}
 \label{FigEth}
\end{figure}

\indent\indent Similarly to ${\rm C_2H_2}$, we compare in Fig.~\ref{FigEth} the ISO-SWS spectrum in the ${\rm C_2H_6}$ $\nu_9$ band with synthetic spectra calculated with three distinct ${\rm C_2H_6}$ vertical profiles (Fig.~\ref{FigPro}). Each of the profiles was designed to reproduce the observed emission in the ${\rm ^{R}Q_{0}}$ multiplet at 12.16 $\mu{\rm m}$, which probes the 1-mbar pressure level. The rest of the $\nu_9$ band consists of weaker multiplets, which sound deeper levels extending from 1 to 10 mbar. This combination of strong and weak multiplets makes this band sensitive to the ${\rm C_2H_6}$ vertical distribution. In addition, the pseudo-continuum level in between the ${\rm C_2H_6}$ emissions is also sensitive to the ${\rm C_2H_6}$ abundance in the lower stratosphere.

Our best-fit model (Fig.~\ref{FigPro}, solid line) has a slope $-d\ln q / d\ln P=0.6$, but the steep-slope model (Fig.~\ref{FigPro}, dotted line) is also marginally compatible with the observations. This results in a relatively large uncertainty on the slope determination: $q=(1.0\pm0.2)\times10^{-5}$ at 1 mbar, and $q=(2.6^{+0.5}_{-0.6})\times10^{-6}$ at 10 mbar, giving $-d\ln q / d\ln P=0.6\pm0.2$. An additional error of 25\% on $q$ comes from the uncertainty on the temperature. Again, it should not affect the retrieved slope.

\subsection{Methylacetylene and diacetylene}

\begin{figure}
 \resizebox{\hsize}{!}{\includegraphics{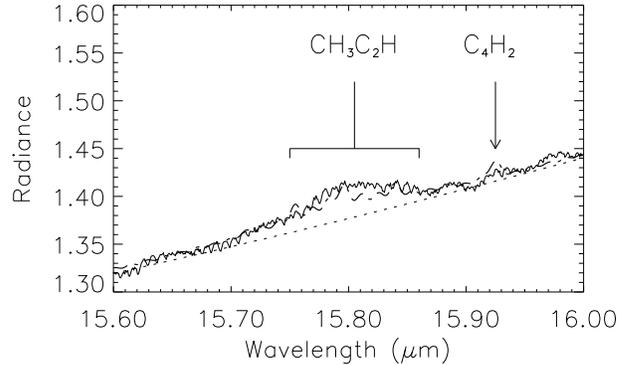}}
 \caption{Comparison between the ISO-SWS spectrum (solid line) and two synthetic spectra; with (dash-dotted line) and without (dotted line) ${\rm CH_3C_2H}$ and ${\rm C_4H_2}$ opacities.}
 \label{FigMet}
\end{figure}

\indent\indent The ISO-SWS spectrum exhibits a broad emission near 15.8 $\mu{\rm m}$, which coincides with the $\nu_9$ band of ${\rm CH_3C_2H}$ (Fig.~\ref{FigMet}). Since the ${\rm CH_3C_2H}$ lines are optically thin, no information on the vertical profile can be derived. Using a vertical profile similar to that calculated by Gladstone et al.\ (\cite{Gladstone96}), we found a column density of $(1.5\pm0.4)\times10^{15}$ molecule\,cm$^{-2}$. The synthetic spectrum exhibits small-scale structures which are not seen in the observations. This mismatch is propably due to an imperfect data reduction. Indeed, the respective frequencies of the fringes and of the ${\rm CH_3C_2H}$ features lie close to each other. It is therefore very difficult to fully remove the former without altering the latter. On the contrary, the broad emission is a low frequency signal and is therefore left unaffected by the frequency filtering. This explanation is admittedly not entirely satisfactory, but given the wavelength match of the emission with the $\nu_9$ mode of ${\rm CH_3C_2H}$, and in the absence of any other plausible candidates, we regard the detection of methylacetylene as unambiguous.

The $\nu_8$ band of ${\rm C_4H_2}$ is not detected at 15.92 $\mu{\rm m}$. We inferred an upper limit of the ${\rm C_4H_2}$ column density of $7\times10^{13}$ molecule\,cm$^{-2}$, using the Gladstone et al.\ (\cite{Gladstone96}) profile.

\section{Discussion}

\indent\indent B\'ezard et al.\ (\cite{Bezard95}) first showed from 13.4-$\mu{\rm m}$ high-resolution spectroscopy that the ${\rm C_2H_2}$ mixing ratio increases with height in the stratosphere, and that most of the acetylene is concentrated above the $\sim$0.5-mbar level. Their distribution has a mixing ratio of about $7\times10^{-7}$ at 0.3 mbar, in reasonable agreement with our results. More recently, B\'etremieux \& Yelle (\cite{Betremieux99}), using UV observations, found an average ${\rm C_2H_2}$ mixing ratio in the 20--60 mbar range of $1.5\times10^{-8}$, consistent with our value of $1.9\times10^{-8}$, extrapolated to this pressure range. Using height-dependent mixing ratio profiles to analyse high-resolution infrared observations, Sada et al.\ (\cite{Sada98}) derived mixing ratios of $3.9^{+1.9}_{-1.3}\times10^{-6}$ for ${\rm C_2H_6}$ at 5 mbar and $2.3\pm0.5\times10^{-8}$ for ${\rm C_2H_2}$ at 8 mbar. While the former value exactly agrees with our results, the latter is almost 3 times less than that extrapolated downwards from our ${\rm C_2H_2}$ profile. Also from infrared observations, Noll et al.\ (\cite{Noll86}), using a slope of $-d\ln q / d\ln P=0.3$, found a ${\rm C_2H_6}$ mixing ratio of $7.5\times10^{-6}$ at 1 mbar, which compares reasonably well with our measurement at the same pressure level.

We also compared our retrieved mixing ratios with those calculated in the photochemical model of Gladstone et al.\ (\cite{Gladstone96}). For ${\rm C_2H_6}$, our results are in excellent agreement with their model both at 1 and 10 mbar (Fig.~\ref{FigPro}). For ${\rm C_2H_2}$, while the agreement is good at 4 mbar, their mixing ratio at 0.3 mbar is higher than ours by a factor of 4. Our ${\rm C_2H_2}$ slope is then significantly lower than theirs ($-d\ln q / d\ln P=0.8\pm0.1$ vs.\ $-d\ln q / d\ln P=1.2$). Note that our derived ${\rm C_2H_2}$ slope is still steeper than that of ${\rm C_2H_6}$ ($-d\ln q / d\ln P=0.6\pm0.2$), as expected. Indeed, ${\rm C_2H_6}$, being less subject to photolysis, has a longer lifetime in the jovian stratosphere than ${\rm C_2H_2}$ (Gladstone et al.\ \cite{Gladstone96}, their Fig.~6).

Hydrocarbons are formed from the photolysis of ${\rm CH_4}$ around the homopause. Small-scale turbulence, parameterized in a photochemical model by the eddy diffusion coefficient ($K$), transports them downwards in the stratosphere. This process is approximately modelled for long-lived products by the equation $K(z)n(z){\rm d}q/{\rm d}z = P(z)$, where $n(z)$ is the number density at altitude $z$ and $P(z)$ the vertically integrated net production rate above altitude $z$. A first hypothesis would attribute the difference between the observed and calculated ${\rm C_2H_2}$ abundances to an overestimation of the ${\rm C_2H_2}$ production rate $P(z)$ in the Gladstone et al.\ model. Our two abundance measurements at 4 and 0.3 mbar allow us to calculate the mean ${\rm d}q/{\rm d}z$ over this pressure range. Comparing with the value of ${\rm d}q/{\rm d}z$ predicted by Gladstone et al.\ at 1 mbar, we found that $P(z)$ should be decreased by a factor of $\sim$3.

A second explanation would be that Gladstone et al.\ underestimated the eddy diffusion coefficient $K$ by a factor of $\sim$3 around 1 mbar. This underestimation at 1 mbar should also apply to the level and eddy diffusion coefficient ($K_H$) at the homopause. However, it is difficult to directly quantify the changes on the homopause parameters, because of the strong coupling between production rates and homopause level. Simply, we note that our analysis could imply an increase in $K_H$. It is consistent with Drossart et al.\ (\cite{Drossart99}), who, analysing ${\rm CH_4}$ fluorescence, found $K_H=(7\pm1)\times10^6$ ${\rm cm^{-2}\,s^{-1}}$, higher than the value of 1.4$\times10^6$ assumed in the Gladstone et al.\ model. In this case, the agreement on the ${\rm C_2H_6}$ slope would also imply that the ${\rm C_2H_6}$ production rate has been underestimated by Gladstone et al.\ (\cite{Gladstone96}). In fact, the most direct conclusion of our measurements is that Gladstone et al.\ have overestimated the ${\rm C_2H_2}$/${\rm C_2H_6}$ production rate ratio.

The ISO-SWS spectrum enables the first detection of ${\rm CH_3C_2H}$ in the equatorial region of Jupiter. We retrieved a column density of $(1.5\pm0.4)\times10^{15}$ molecule\,cm$^{-2}$. Kim et al.\ (\cite{Kim85}) had previously detected ${\rm CH_3C_2H}$ in the north polar auroral zone of Jupiter, and had retrieved a column density of $(2.8^{+2.4}_{-1.1})\times10^{16}$ molecule\,cm$^{-2}$. At least a large part of the difference is explained by different modelling assumptions. Kim et al.\ (\cite{Kim85}) assumed a uniform vertical distribution throughout the stratosphere and used a temperature profile for the auroral region which is now known to be incorrect (Drossart et al.\ \cite{Drossart93}). For ${\rm C_4H_2}$, we found only an upper limit of $7\times10^{13}$ molecule\,cm$^{-2}$, 65 times lower than the stratospheric column density predicted by Gladstone et al.\ (4.5$\times$10$^{15}$ molecule\,cm$^{-2}$). This is consistent with their overestimate of ${\rm C_2H_2}$ production, since the production of ${\rm C_4H_2}$ is essentially quadratically dependent on the abundance of ${\rm C_2H_2}$. As our ${\rm CH_3C_2H}$ column density is 3.5 times larger than calculated by Gladstone et al., we derived a ${\rm CH_3C_2H}$/${\rm C_4H_2}$ ratio larger than 20, while they found a ratio of $\sim2$. This discrepancy is all the more remarkable as, in the case of Saturn, where both ${\rm CH_3C_2H}$ and ${\rm C_4H_2}$ are detected (de Graauw et al.\ \cite{deGraauw97}), the photochemical model of Moses et al.\ (\cite{Moses99}) gives, as observed, a ${\rm CH_3C_2H}$/${\rm C_4H_2}$ ratio of about 10. This stresses that the C$_3$ and C$_4$ chemistry in Jupiter should be reassessed with a complete photochemical model of the jovian stratosphere.

\begin{acknowledgements}
This study is based on observations with ISO, an ESA project with instruments funded by ESA Member States (especially the principal investigators countries: France, Germany, the Netherlands and the United Kingdom), and with participation of ISAS and NASA.
\end{acknowledgements}

\end{document}